\documentclass[11pt, a4paper]{article} 
\pdfoutput=1

\usepackage[ansinew]{inputenc}
\usepackage{amsmath, amssymb, graphics, amsthm}
\usepackage{epsfig}
\usepackage{color, xcolor}
\usepackage{fancyhdr} 
\usepackage{manfnt}
\usepackage[T1]{fontenc}
\usepackage{hyperref}

\usepackage[normalem]{ulem}

\newcommand{\vev}[1]{\left\langle #1 \right\rangle}

\makeatletter
\@addtoreset{equation}{section}
\makeatother

\newcommand{\be}{\begin{equation}}
\newcommand{\ee}{\end{equation}}
\newcommand{\ba}{\begin{eqnarray}}
\newcommand{\ea}{\end{eqnarray}}

\def\pb#1{\rlap{\lower1.5ex\hbox{$\longleftarrow$}}{#1}}
\def\dpb#1{\rlap{\lower1.5ex\hbox{$\Longleftarrow$}}{#1}}
\def\spb#1{\rlap{\lower1.0ex\hbox{$\leftarrow$}}{#1}}
\def\sdpb#1{\rlap{\lower1.0ex\hbox{$\Leftarrow$}}{#1}}

\newcommand{\amin}{a_\text{ext}}

\newcommand{\lbdy}{\mathcal L_\text{bdy}}

\newcommand{\lren}{L_{\text{ren}}}

\newcommand{\tlau}{\left( \frac{t^2 + \lambda^2}{t_*^2 + \lambda^2} \right)}

\title{{\sf Holographic signatures of resolved cosmological singularities}}
\author{
{\sf N. Bodendorfer\thanks{{\sf 
norbert.bodendorfer@lmu.de}}, A. Sch\"afer\thanks{{\sf 
andreas.schaefer@physik.uni-regensburg.de}}, J. Schliemann\thanks{{\sf john.schliemann@physik.uni-regensburg.de}}}\\
\\
{\sf Institute for Theoretical Physics, University of Regensburg, } \\ {\sf 93040 Regensburg, Germany  }}
\date{{\small\sf \today}}

\begin{document} 

\maketitle

{\sf

\begin{abstract}

The classical gravity approximation is often employed in AdS/CFT to study the dual field theory, as it allows for many computations. A drawback is however the generic presence of singularities in classical gravity, which limits the applicability of AdS/CFT to regimes where the singularities are avoided by bulk probes, or some other form of regularisation is applicable. At the same time, quantum gravity is expected to resolve those singularities and thus to extend the range of applicability of AdS/CFT also in classically singular regimes. This paper exemplifies such a computation. We use an effective quantum corrected Kasner-AdS metric inspired by results from non-perturbative canonical quantum gravity to compute the 2-point correlator in the geodesic approximation for a negative Kasner exponent. The correlator derived in the classical gravity approximation has previously been shown to contain a pole at finite distance as a signature of the singularity. Using the quantum corrected metric, we show explicitly how the pole is resolved and that a new subdominant long-distance contribution to the correlator emerges, caused by geodesics passing arbitrarily close to the resolved classical singularity. In order to compute analytically in this paper, two key simplifications in the quantum corrected metric are necessary. They are lifted in a companion paper using numerical techniques, leading to the same qualitative results.

\end{abstract}

}

\section{Introduction}

The classical supergravity approximation is extensively used in the AdS/CFT correspondence \cite{MaldacenaTheLargeN, WittenAntiDeSitter, GubserGaugeTheoryCorrelators} (see \cite{AmmonGaugeGravityDuality} for a textbook) for the simple reason that it allows to perform explicit computations in many situations. It corresponds to taking the limit of large 't Hooft coupling and large central charge in the dual field theory, and thus unfortunately not to the regime that one is mainly interested for experiments. The most interesting regime of finite 't Hooft coupling and  finite central charge is that of full string theory, where explicit computations are usually out of reach. In particular, the currently best formulation of non-perturbative string theory, which is required for such computations, is the AdS/CFT correspondence itself, i.e. non-perturbative string theory is defined via its dual CFT. 

Meanwhile, one generically encounters singularities in the classical supergravity approximation which can be reached by bulk probes, see e.g. \cite{EngelhardtHolographicSignaturesOf, EngelhardtFurtherHolographicInvestigations} and references therein. In particular, such singularities can lead to strange behaviour in the dual CFT, such as poles in the two-point correlator at finite distance \cite{EngelhardtHolographicSignaturesOf, EngelhardtFurtherHolographicInvestigations}. This situation is unsatisfactory, as the general belief in the field is that singularities should be resolved by quantum gravity effects\footnote{For example, notable progress has been made in understanding singularities in the tensionless limit of string theory which is described by higher-spin theories \cite{AmmonSpacetimeGeometryIn, KrishnanDesingularizationOfThe, CrapsLowTensionStrings, KiranStringsVsSpins}.}. 

In this situation, it is natural to ask how singularities are resolved in other approaches to quantum gravity. Given a particular quantum corrected metric where the singularity is resolved, one can use this metric in the AdS/CFT correspondence and study its implications for the dual field theory. The main conceptual question here is whether such a metric is a good approximation to what would happen in string theory. Note however that we already defined non-perturbative string theory via AdS/CFT, so that one can ask in first approximation: ``Does the quantum corrected metric give sensible results for the dual CFT?''.

In this paper, we will give a prototype calculation where the answer to this question is ``yes''. We use a quantum corrected metric inspired by results from loop quantum gravity, where the singularity is resolved and thereby extend the results of \cite{EngelhardtHolographicSignaturesOf, EngelhardtFurtherHolographicInvestigations} obtained at the classical level. For this metric, we use the geodesic approximation to compute the two-point correlator of two heavy scalar operators. 
It turns out that a pole in the correlator which is present in the classical theory is resolved for the quantum corrected metric. Furthermore, a new subleading long-distance contribution to the correlator is found, whose functional dependence on the spatial separation agrees with its standard short-distance behaviour. 

Whether these properties of the dual CFT are indeed those of $\mathcal N = 4$ super Yang Mills theory, the conjectured dual of type IIB string theory, remains open. A possible strategy to answer this question is to resort to lattice simulations, see e.g. \cite{HanadaWhatLatticeTheorists}. In fact, this route seems to be the most promising one if one wants to establish whether a given theory of quantum gravity is a good approximation to string theory.  \\

This paper is organised as follows:\\
We recall results obtained in the classical gravity approximation in section \ref{sec:Class}. The main part of the paper is section \ref{sec:Quant}, where the geodesic equation is solved for the quantum corrected metric and the two-point correlator is extracted. We provide some comments in section \ref{sec:Comments} and conclude in section \ref{sec:Conclusion}.

\section{CFT correlators from classical Kasner metrics} \label{sec:Class}

In this section, we will recall the results of a recent series of papers \cite{EngelhardtHolographicSignaturesOf, EngelhardtFurtherHolographicInvestigations} on CFT signatures of cosmological bulk singularities, see also \cite{HertogTowardsABig, HertogHolographicDescriptionOf, DasTimeDependentCosmologies, TurokFromBigCrunch, DasCosmologiesWithNull, CrapsQuantumResolutionOf, AwadGaugeTheoryDuals, AwadGaugeTheoriesWith, BarbonAdSCrunchesCFT, SmolkinDualDescriptionOf} for earlier work. In the next section, we will generalise these results to a 1-parameter family of quantum corrected metrics labelled by $\lambda \geq 0$, from which the classical result can be obtained in the limit $\lambda \rightarrow 0$. Further work generalising \cite{EngelhardtHolographicSignaturesOf, EngelhardtFurtherHolographicInvestigations} to other classical cosmological spacetimes was done in \cite{ChatterjeeNonVacuumAdS}. 

The setup of \cite{EngelhardtHolographicSignaturesOf, EngelhardtFurtherHolographicInvestigations} is to consider Kasner-AdS bulk spacetimes, which are given by the metric
\be
	ds_5^2 = \frac{L^2}{z^2} \left( dz^2+ds_4^2(t)  \right), ~~~~~~ ds_4^2(t) = -dt^2 + \sum_{i=1}^3 t^{2p_i} dx_i^2 \text{.}
\ee
The $p_i$s obey the vacuum Kasner conditions $p_1 + p_2 + p_3 = 1 =  p_1^2 + p_2^2 + p_3^2$, so that $ds_4^2$ is a solution to the four-dimensional vacuum Einstein equations without cosmological constant. $ds_5^2$ solves the five-dimensional vacuum Einstein equations with negative cosmological constant $\Lambda = - 10/L^2$. We will set $L=1$ from now on and only restore it in the qualitative discussion in the comment section \ref{sec:Comments}. 

Following the AdS/CFT dictionary, this bulk system is equivalent to $\mathcal N = 4$ Super Yang Mills theory on a Kasner background. In particular, the two-point correlators of heavy ($m \gg 1$) scalar operators can be computed via the geodesic approximation \cite{BalasubramanianHolographicParticleDetection}
\be
	\vev{\mathcal{O}(x) \mathcal{O}(-x)} \sim \exp(- \Delta \lren ), ~~~~~~ \Delta = d/2 + \sqrt{d^2/4+m^2} \label{eq:GeodesicApprox} \text{,}
\ee
where $\lren$ is the renormalised length of a spacelike geodesic connecting the boundary points $(t_0,x)$ and $(t_0,-x)$. In case of multiple geodesics satisfying given boundary data, one has to sum over the individual contributions. Complex solutions also have to be taken into account.

The main motivation of \cite{EngelhardtHolographicSignaturesOf, EngelhardtFurtherHolographicInvestigations} was to study CFT signatures of the bulk singularity. For this, the bulk geodesics were computed as a function of their turning time $t_*$. If we consider the x-separation in a direction where $p_i<0$, the geodesics are curved towards the singularity and $t_* < t_0$ for real solutions, and the other way around for $p_i > 0$. In the limit $t_* \rightarrow 0$, the geodesic becomes null and its tip approaches the bulk singularity. The CFT signature of this is a pole in the two-point correlator at the cosmological horizon scale. 

This pole signals that the state in the dual field theory description of the Kasner-AdS metric cannot be normalisable \cite{EngelhardtFurtherHolographicInvestigations}. It was then argued that quantum effects might smoothen out the pole, however no example or mechanism for this was given. 
Complex geodesics were also taken into account and it was found that they need to be included to ensure smoothness of the two-point correlator. Solutions where the geodesic crossed the singularity were discarded, as the geodesic approximation is not justified in such a case. For a direction with $p_i=-1/4$, this means that only one to two out of five possible solutions to the geodesic equation can be taken into account reliably \cite{EngelhardtHolographicSignaturesOf}.

\section{Improved CFT correlators from quantum gravity} \label{sec:Quant}

\subsection{Motivation for the choice of metric}

We have recalled in the previous section that the singularity occurring in the classical Kasner metric leads to a pole in the two-point correlator of the dual CFT at horizon scale. It was already discussed in \cite{EngelhardtFurtherHolographicInvestigations} that quantum gravity effects might smoothen out this pole and render the the two-point correlator finite at non-vanishing spatial separation. In this section, we want to give an explicit example for this. Our strategy is to consider effective spacetimes emerging from quantum gravity and to continue using the geodesic approximation therein. This should be justified in a region where the theory behaves like a classical gravitational theory with higher curvature corrections (see however the comment section \ref{sec:Comments}). A similar strategy was already used in \cite{AshtekarCovariantEntropyBound} to show that the covariant holographic entropy bound \cite{BoussoACovariantEntropy} can be satisfied in presence of a singularity that has been resolved by quantum effects.

Since the 5d metric
\be
	ds_5^2 = \frac{1}{z^2} \left( dz^2+ds_4^2(t)  \right) \label{eq:Metric4d5d}
\ee
is singular only in its four-dimensional part and the 5d-Einstein equations with negative cosmological constant imply that $ds_4^2$ is Ricci-flat, we can look for a quantum corrected version of $ds_4^2$ using 4d quantum gravity with vanishing cosmological constant. This means that we keep the components of the metric in $z$-direction classical since no Planck regime curvatures are associated with them (we take the cosmological constant to be small enough for this to be true), and only quantise the components orthogonal to the $z$-direction. A more subtle but important choice in the quantisation prescription determining the magnitude of the quantum effects in the 4d/5d interplay is discussed in section \ref{sec:Comments}. The choice in this matter made here is motivated by allowing for analytic computations in the following, and is not the most natural one. A companion paper \cite{BodendorferHolographicSignaturesOf2} will deal with improved metrics using numerical techniques.
Here, we consider the diagonal 4d-metric 
\be
	ds_4^2 = - dt^2 + a(t)^2 d x^2 + \ldots, ~~~~~~ a(t) = \frac{\amin}{\lambda^{p}} \left(t^2 + \lambda^2 \right)^{p/2} \label{eq:QuantumMetric}
\ee
as a quantum corrected Kasner metric, where $\ldots$ refers to the other spatial directions which (may) have different Kasner exponents. $\amin$ denotes the (extremal) value that the scale factor takes at $t=0$. $\lambda$ measures the scale at which quantum gravity effects become important, i.e. it contains $\hbar$. For $\lambda>0$, the classical singularity is resolved. In the double scaling limit $\lambda \rightarrow 0$ with $\amin / \lambda^{p} = 1 $, the classical Kasner solution with $a(t) =  t^{p}$ is obtained. The time of bounce (extremal scale factor) has been chosen to be $t=0$, but can be set to any time $t_0$ by the replacement $t \mapsto t-t_0$ in all formulas. 

Our motivation for this form of the quantum corrected Kasner metric stems from loop quantum gravity\footnote{Similar metrics can in principle be motivated by any modification of general relativity leading to bouncing solutions.}. Here, the best studied scenario is spatially flat homogeneous and isotropic cosmology sourced by a massless scalar field (an equation of state with $\omega = 1$), where all three Kasner exponents are given by $p=1/3$. In this case, \eqref{eq:QuantumMetric} can be derived as an exact solution of a minisuperspace quantisation \cite{AshtekarRobustnessOfKey} (see also \cite{BodendorferAnElementaryIntroduction}), which can be embedded into a full theory setting \cite{BodendorferStateRefinementsAnd}, including an explicit continuum limit of the quantum geometry \cite{BVI}. Several other works also strengthen these results\footnote{A very similar form of the dynamics has been derived also using group field theory \cite{OritiEmergentFriedmannDynamics} and using improved regularisations in the canonical theory \cite{AlesciImprovedRegularizationFrom}. Similar results have also been reported in \cite{ChamseddineResolvingCosmologicalSingularities}, however with the aim to construct a classical gravitational theory with a build-in limiting curvature.} and phenomenological investigations based on them are being undertaken \cite{AshtekarQuantumGravityExtension, AshtekarLoopQuantumCosmologyFrom, BollietObservationalExclusionOf}. 
The numerical value of $\lambda$ is a quantisation ambiguity in the theory that can be directly related to the choice of Barbero-Immirzi parameter \cite{AshtekarQuantumNatureOf}, and is expected to be at the order of the Planck length.

In the general non-isotropic case, no analytic solution is known. Using effective equations derived from expectation values of the minisuperspace Hamiltonian constraint operator, the quantum dynamics have been investigated numerically in \cite{GuptQuantumGravitationalKasner}. It was found that the singularity gets resolved and is replaced by a smooth transition between Kasner universes. The detailed behaviour of the solutions is more intricate than that of \eqref{eq:QuantumMetric}. In particular, Kasner exponents may smoothly change during the transition. While positive Kasner exponents may transition into other positive Kasner exponents, negative exponents always change into positive ones. This is in stark contrast to \eqref{eq:QuantumMetric}, which features a negative to negative transition. 

In this paper, we still chose to work with \eqref{eq:QuantumMetric}, for the simple reason that it allows for analytic computations in the following. As said before, we will tackle the issue of Kasner transitions and a proper setting of the 5d Planck scale in a companion paper \cite{BodendorferHolographicSignaturesOf2}. It will turn out that the qualitative form of the two-point boundary correlator derived from \eqref{eq:QuantumMetric} is insensitive to these changes: its finite distance pole is resolved in all cases in a qualitatively similar way. Still, even the improved forms of the metrics discussed in \cite{BodendorferHolographicSignaturesOf2} will have some insufficiencies.

Therefore, we see our computation here and in \cite{BodendorferHolographicSignaturesOf2} only as a proof of principle that non-perturbative quantum gravity can give a significant improvement over the classical gravity approximation in AdS/CFT. 
Further research aimed at obtaining better effective metrics is certainly necessary.

\subsection{Solution of the geodesic equation}

\subsubsection{Coordinate parametrisation}

The non-vanishing Christoffel symbols computed from \eqref{eq:QuantumMetric} are
\begin{align}
	\Gamma^t_{xx}&= p \frac{\amin^2}{\lambda^{2p}}  {t \left(t^2 + \lambda^2\right)^{-1+p}}, ~~~ & \Gamma^t_{zt}& = -\frac{1}{z} , ~~~ & \\
	\Gamma^x_{xt} &= \frac{p t}{\left(t^2+\lambda^2\right)}, 	&\Gamma^x_{zx} & = - \frac{1}{z}, &\\
	\Gamma^z_{tt} & = - \frac{1}{z}, & \Gamma^z_{xx} &= \frac{\amin^2}{\lambda^{2p}} \, \frac{ \left(t^2 + \lambda^2 \right)^{p}}{z}, &	\Gamma^z_{zz} &= - \frac{1}{z} \text{,}
\end{align}
as well as Christoffel symbols involving the other spatial directions. In the following, we will use greek letters $\alpha, \beta, \gamma, \ldots$ from the beginning of the alphabet to denote tensor indices in the 5d spacetime, and greek letters $\mu, \nu, \rho, \ldots$ from the middle of the alphabet for 4d tensor indices for the 4d cosmological spacetime embedded into 5d AdS.

Following \cite{EngelhardtFurtherHolographicInvestigations}, it is most convenient to solve the geodesic equation when parametrised with the time coordinate $t$. In this case, the geodesic equation is given by
\be
	\ddot{x}^{\alpha} + \Gamma^\alpha_{\beta \gamma} \dot x^\beta \dot x^\gamma - \Gamma^t_{\beta \gamma} \dot x^\beta \dot x^\gamma \dot x^\alpha =0 \text{.}
\ee
The equation for the $x$-component reads
\be
	\ddot x +  \frac{2 p t}{t^2 + \lambda^2} \dot x -p  \frac{\amin^2}{\lambda^{2p}}  t \left( t^2 + \lambda^2 \right)^{-1+p} \dot x^3 = 0
\ee
and is solved by
\be
	\dot x(t) = \pm \frac{ 1}{\sqrt{\frac{\amin^2}{\lambda^{2p}} \left( t^2 + \lambda^2 \right)^{p} + c (t^2 + \lambda^2)^{2p}}   } \text{.}
\ee
Integration w.r.t. $t$ then gives $x(t)$, however no explicit antiderivative seems to be known. It is convenient to introduce the parameter $t_*$ satisfying $t_*^2 + \lambda^2 = \frac{\amin^{2/p}}{|c|^{1/p} \lambda^2}$ and to abbreviate $\tau = \tlau$. For $t = t_*$, we have $\dot x ^{-1} = \frac{dt}{dx}= 0$, i.e. we are at the turning point of the geodesic. This also implies that $c < 0$ for the geodesic to exist. It follows that
\be
	\dot x(t) = \pm \frac{\lambda^{p}}{\amin} \left(t_*^2 + \lambda^2\right)^{-p/2} \frac{ 1}{\sqrt{ \tau^{p} \left( 1-\tau^{p} \right) }}   \text{.} \label{eq:xdot}
\ee 
We note that $\tau^p \leq 1$ both for $p>0$ and $p<0$, if geodesics for $p>0$ are curved away from the classical singularity ($t_*>t_0$), while geodesics with $p<0$ are curved towards it ($t_*<t_0$). We restrict us here to these cases to ensure reality.

The equation for the $z$-component with the definition $v := z \dot z$ reads
\begin{align}
	\dot v &= 1 - \dot x^2 \frac{\amin^2}{\lambda^{2p}} (t^2 + \lambda^2)^{p} + p v \dot x^2 \frac{\amin^2}{\lambda^{2p}} \, t \left(t^2 + \lambda^2 \right)^{-1+p} \nonumber \\
		& = 1 - \frac{1}{1-\tau^{p}} + p v \frac{t}{t_*^2 + \lambda^2} \frac{1}{\tau-\tau^{1+p}} \text{.}
\end{align}
This equation can be solved by the ``variation of parameters'' method, giving
\be
		v(t) = z(t) \dot z(t)= c_3 \frac{\tau^{p/2}}{\sqrt{1-\tau^{p}}}  - \frac{\tau^{p/2}}{\sqrt{1-\tau^{p}}} \int_{t_*}^t dt' \frac{{\tau'}^{p/2}}{\sqrt{1-{\tau'}^{p}}} \label{eq:v}
\ee	
Smoothness of the geodesic at the turning point demands $\frac{dz}{dx}(t_*) =  \frac{\dot z}{  \dot x} (t_*)=0$, which implies $c_3 = 0$. $z(t)$ can be obtained now by integrating \eqref{eq:v}:
\be
	z(t) = \sqrt{ z(t_*)^2 - \left( \int_{t_*}^t dt'  \frac{{\tau'}^{p/2}}{\sqrt{1-{\tau'}^{p}}} \right)^2  } \text{.} \label{eq:ZSolution}
\ee
We see that $z(t) < z(t_*)$ for $t \neq t_*$, in accordance with the geometric properties of the geodesic. Furthermore, it can be checked that the geodesic is spacelike as long as $t_*>0$. 

There are two non-trivial parameters defining our geodesic: $t_*$ and $z(t_*)$. They determine the time $t_0$ at which $z = 0$, i.e. the time at which the geodesic intersects the boundary, as well as the boundary separation in $x$-direction. The additional integration constant appearing in $x(t) = \int^t dt' \dot x(t')$ will be fixed so that $x(t_*) = 0$.
Alternatively, we may choose to specify $t_*$ and $t_0$. Then, $z(t_*) = \left| \int_{t_*}^{t_0} dt'  \frac{{\tau'}^{p/2}}{\sqrt{1-{\tau'}^{p}}}  \right|$ and the {\it proper} separation of the geodesic endpoints in $x$-direction is $\lbdy = 2\, a(t_0) \left| \int_{t_*}^{t_0} dt' \dot x(t') \right| = 2 a(t_0) x(t_0)$.

\subsubsection{Affine parametrisation for $z$}

While the geodesic equation parametrised w.r.t. the time coordinate $t$ could be solved completely, it has the drawback that the final result is hard to handle in the context of holographic renormalisation, i.e. the subtracting its diverging part, see below. In order to circumvent this problem, we will now derive a solution $z(s)$ parametrised w.r.t. to the geodesic length $s$ from the affinely parametrised geodesic equation. 

The equation for $z(s)$ reads
\be
	{z''}- \frac{1}{z} {t'}^2+\left( \frac{\amin}{\lambda^p}\right)^2 \frac{\left(t^2 + \lambda^2\right)^p}{z} {x'}^2 - \frac{1}{z}  {z'}^2 = 0 \text{.} \label{eq:AffineZEq}
\ee
Using $g_{\alpha \beta} \frac{dx^\alpha}{ds}\frac{dx^\beta}{ds} = 1$, \eqref{eq:AffineZEq} simplifies to 
\be
	z''- \frac{2}{z} {z'}^2+ z=0 
\ee
and can be solved as
\be
	z(s) = \frac{z(t_*)}{\cosh(s-s_0)} \text{.} \label{eq:ZAffineSol}
\ee
We set $s_0 = 0$ to start counting the proper distance form the turning point of the geodesic. Explicit solutions $x(s)$ and $t(s)$ are not needed in the following.

\subsection{Renormalised geodesic length}

Equation \eqref{eq:ZAffineSol} clearly shows that the length of our geodesics always diverge. We are therefore in need for a renormalisation procedure. For this, we let the geodesic end not at $z=0$, but at $z=\epsilon$, and subtract the occurring divergence. In the limit $\epsilon\rightarrow 0$, we have 
\be
	\pm s(z = \epsilon) =  \log \left(2 z(t_*) \right) - \log(\epsilon) \text{.}
\ee
Subtracting the divergent piece $- \log(\epsilon)$ can be understood as the effect of a conformal transformation that removes the conformal factor $1/z^2$ in $ds^2_5$ from the boundary metric, leading to the correct 2-point correlator expression for the boundary metric $ds_4^2$ \cite{BalasubramanianHolographicParticleDetection}. This leaves us with the renormalised geodesic length
\be
	\lren = 2 \log (2 z(t_*)). 
\ee
As a cross-check, we can specialise to $p=-1/2$ and take the classical limit $\lambda \rightarrow 0$ to obtain the result $\lren = \log \left( 16 t_* (1-t_*) \right)= \log \left( - 16 c (1+c) \right)$ derived before equation (5.3) of \cite{EngelhardtFurtherHolographicInvestigations}.

\subsection{Contributions to the two-point correlator from real geodesics}

\subsubsection{Short distance behaviour}

To compute the short distance behaviour, we define $\epsilon = t-t_*$ and compute for $p<0$
\begin{eqnarray}
	z(t_*) &=& \int_{0}^{t_0-t_*} d\epsilon \,   \frac{{\tau}^{p/2}}{\sqrt{1-{\tau}^{p}}} = \int_{0}^{t_0-t_*} d\epsilon \left( \sqrt{ \frac{t_*^2+\lambda^2}{-2 p t_* \epsilon} } + \mathcal{O} (\epsilon^{1/2})\right) \nonumber \\
	&=& \sqrt{2} \sqrt{ \frac{t_*^2+\lambda^2}{-p t_*} } \sqrt{t_0-t_*} + \mathcal{O}\left( (t_0-t_*)^{3/2} \right) \text{.}
\end{eqnarray}
A similar calculation can be done for $x(t_0)$. It follows that

\be
	\vev{\mathcal{O}(x) \mathcal{O}(-x)}   \stackrel{x \rightarrow 0}{\sim} \left( 2x \frac{\amin}{\lambda^p} (t^2+\lambda^2)^{p/2} \right)^{- 2 \Delta} = \left( \lbdy   \right)^{- 2 \Delta} \text{.}
\ee
The same result also follows for $p>0$, for which only the integration borders have to be switched due to $t \leq t_*$. 
We observe that the short distance behaviour remains invariant in the classical limit, which is in agreement with the results of \cite{EngelhardtFurtherHolographicInvestigations}.

\subsubsection{Long distance behaviour, $p<0$}

We note that for $t_* \rightarrow 0$, $\lbdy \rightarrow \infty $. Therefore, real geodesics exist for all $\lbdy \geq 0$. Moreover, for large enough $\lbdy$, the (real) geodesic is unique (see figure \ref{fig:PlotXZ}). 
At the same time, we have $z(t_*) \rightarrow \infty$ for $t_* \rightarrow 0$.
In order to study the long distance behaviour of the two-point correlator, we need to express $z(t_*)$ in terms of $\lbdy$ for large $\lbdy$. We compute
\begin{eqnarray}
	\lim_{t_* \rightarrow 0} \frac{2 z(t_*)}{\lbdy} &=& \frac{1}{\left(t^2 + \lambda^2 \right)^{p/2}} \lim_{t_* \rightarrow 0} \frac{  \int_{t_*}^{t_0} dt  \frac{{\tau}^{p/2}}{\sqrt{1-{\tau}^{p}}} }{ \int_{t_*}^{t_0} dt   \left(t_*^2 + \lambda^2\right)^{-p/2} \frac{ 1}{\sqrt{ \tau^{p} \left( 1-\tau^{p} \right) }}} \nonumber \\
		& =&  \frac{1}{\left(1+t^2/ \lambda^2 \right)^{p/2}} \lim_{t_* \rightarrow 0} \frac{  \int_{t_*}^{t_0} dt  \frac{ (\tau^p-1) {\tau}^{-p/2} + \tau^{-p/2}}{\sqrt{1-{\tau}^{p}}} }{ \int_{t_*}^{t_0} dt   \frac{ \tau^{-p/2}}{\sqrt{  \left( 1-\tau^{p} \right) }}} \nonumber \\
		& =&  \frac{1}{\left(1+t^2/ \lambda^2 \right)^{p/2}} \lim_{t_* \rightarrow 0} \left(1- \frac{  \int_{t_*}^{t_0} dt   \sqrt{1-\tau^p} {\tau}^{-p/2}} { \int_{t_*}^{t_0} dt   \frac{ \tau^{-p/2}}{\sqrt{  \left( 1-\tau^{p} \right) }}} \right)  \nonumber \\
		&=& \frac{1}{\left(1+t^2/ \lambda^2 \right)^{p/2}}  = \frac{1}{a(t)} \frac{\amin}{\lambda^p} \lambda^p  \text{.} \label{eq:XZBehaviour}
\end{eqnarray}
Insertion into \eqref{eq:GeodesicApprox} gives
\be
	\vev{\mathcal{O}(x) \mathcal{O}(-x)}  \stackrel{x \rightarrow \infty}{\sim} \left( 2x \frac{\amin}{\lambda^p} \lambda^p \right)^{- 2 \Delta} = \left( \frac{\lambda^2}{t^2+\lambda^2} \right)^{-  p \Delta}  \left( \lbdy  \right)^{- 2 \Delta} \text{.} \label{eq:RealContributionLR}
\ee
We see that for large distances, the two-point correlator falls off as expected in a standard conformal field theory, modulo a time dependent constant. The result vanishes in the classical limit $\lambda \rightarrow 0$, $\amin/\lambda^p = 1$. The dependence of the final result on $\lambda$ is non-analytic (unless $- p \Delta \in \mathbb{N}_0$ in the expression involving $x$), which signals that it is unlikely to obtain it in a perturbative expansion around $\lambda = 0$. 

We note that for $\lambda > 0$, the geodesic does not approach the boundary as $t_*\rightarrow 0$, since $-c \rightarrow  \left( \frac{\amin}{\lambda^p} \right)^2 \frac{1}{\lambda^{2p}}>0$ in this case. $c=0$, leading to a null geodesic on the boundary \cite{EngelhardtFurtherHolographicInvestigations}, is only recovered in the classical limit.

\subsubsection{Long distance behaviour, $p>0$}

For the long distance behaviour, we note that both 
\be
	 z(t_*) = \int_{t_0}^{t_*} dt  \frac{{\tau}^{p/2}}{\sqrt{1-{\tau}^{p}}} ~~~ \text{and} ~~~ x(t_0) =  \int_{t_0}^{t_*} dt  \frac{\lambda^p}{\amin} \frac{{\tau}^{-p/2} (t_*^2+\lambda^2)^{-p/2}}{\sqrt{1-{\tau}^{p}} }
\ee
diverge as $t_* \rightarrow \infty$. Both integrands are finite unless $t=t_*$, which means that all contributions to the integral until some given value of $t = \tilde t$ remain finite. We can choose $\tilde t$ large enough so that the classical limit $\lambda \rightarrow 0$, $\amin/\lambda^p = 1 $ is an excellent approximation. In this limit, the integrals can be explicitly performed using computer algebra. We find that
\be
	\frac{z(t_*)}{x(t_0)} \xrightarrow{t_* \rightarrow \infty} (1-p)  t_*^p, ~~~~~~ z(t_*)  \xrightarrow{t_* \rightarrow \infty} \frac{\sqrt{\pi}  \Gamma(\frac{1}{2}(3+\frac 1p))}{(1+p) \Gamma(1+\frac{1}{2p})}t_*
\ee
It follows that 
\be
	\vev{\mathcal{O}(x) \mathcal{O}(-x)}  \stackrel{x \rightarrow \infty}{\sim} \text{const}(p) \left( 2x  \right)^{- \frac{2 \Delta}{1-p}} = \text{const}(p) \left(\frac{\lambda^2/a_\text{ext}^{2/p}}{t^2+\lambda^2} \right)^{-  \frac{p \Delta}{1-p}} \left(\lbdy  \right)^{- \frac{2 \Delta}{1-p}}  \text{,} \label{eq:LongDistancePg0}
\ee
confirming the conjecture made in \cite{EngelhardtHolographicSignaturesOf} in the context of the classical theory. 
As noted already there, this behaviour disagrees with the short distance one. In addition, we also observe that it disagrees with the real geodesic long distance behaviour for $p<0$. The case $p=1$ is special and discussed in \cite{EngelhardtHolographicSignaturesOf}.

\subsubsection{Intermediate distance behaviour}

		\begin{figure}[!h]
	\centering\includegraphics[trim = 0mm 0mm 0mm 0mm, clip, scale=0.7]{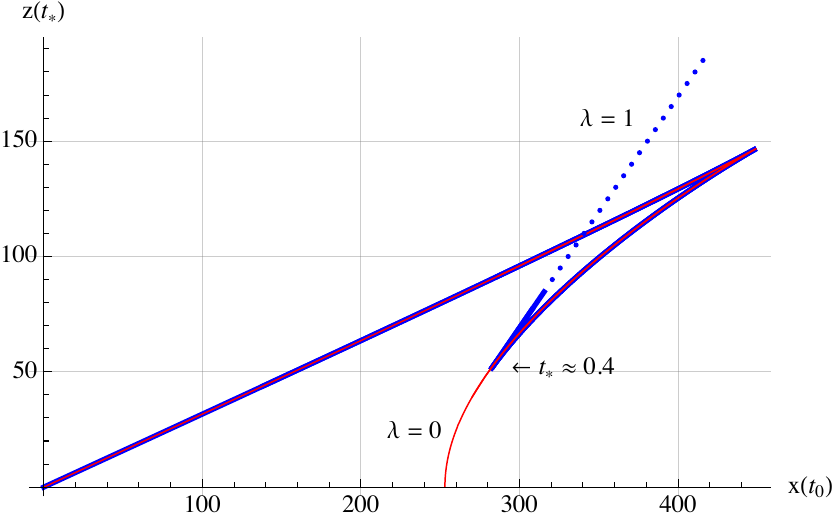} 
		\caption{$z(t_*)$ is plotted against $x(t_0)$ for $\lambda = 1$ (thick blue) and $\lambda = 0$ (thin red), starting from $t_* = t_0 = 100$ at $(0,0)$. The solid blue line was obtained from numerical computations, while the dashed blue line shows the asymptotic behaviour for $t_*\rightarrow 0$, which is hard to probe numerically (the crossover to the blue dashed line is at $ t_* = 1.4 \cdot 10^{-11}t_0$), but has been computed analytically in equation \eqref{eq:XZBehaviour}. In the classical limit (red curve), $x(t_0)$ approaches half the cosmological horizon scale for $t_* \rightarrow 0$, in this case $80\sqrt{10} \approx 253$.
		We note that the same $x(t_0)$-value corresponds to multiple $z(t_*)$ values, which we have to add in the two-point correlator (in addition to complex solutions). We also note that the resolved classical pole is still the dominant (smallest $z(t_*)$) contribution around its $x(t_0)$ value. This behaviour turned out to be generic for several other cases we have tested whenever $t_0 \gg \lambda$. The blue line starts to deviate significantly from the red line around $t_* \approx 0.4$.}
		\label{fig:PlotXZ}
	\end{figure}

			\begin{figure}[!h]
	\centering\includegraphics[trim = 0mm 0mm 0mm 0mm, clip, scale=0.7]{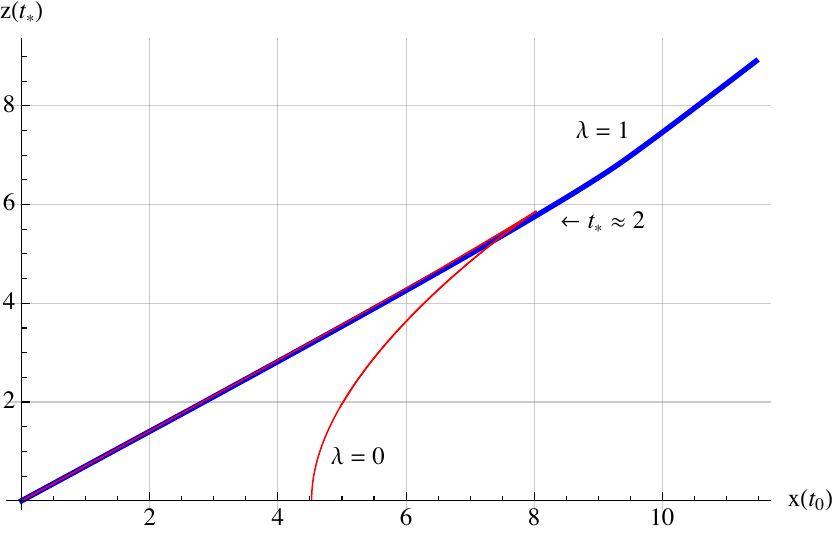} 
		\caption{$z(t_*)$ is plotted against $x(t_0)$ for $\lambda = 1$ (thick blue) and $\lambda = 0$ (thin red), starting from $t_* = t_0 = 4$ at $(0,0)$. The characteristic intermediate scale behaviour shown in figure \ref{fig:PlotXZ} disappears starting around $t_0 \lesssim 5$, i.e. when quantum corrections start to become relevant in the background spacetime of the CFT. We note that the change of slope of the blue curve, here around $x(t_0) = 8.5$, still persists. }
		\label{fig:PlotXZt0}
	\end{figure}

In order to investigate the intermediate distance behaviour of the two-point correlater, we plot $z(t_*)$ vs. $x(t_0)$ for the case $\amin = 1$, $p=-1/4$, for the two values $\lambda = 1$ corresponding to the quantum theory and $\lambda = 0$ corresponding to the classical theory in figure \ref{fig:PlotXZ} for and $t_0=100$ and figure \ref{fig:PlotXZt0} for $t_0=4$. $t_* = t_0$ corresponds to the point $(0,0)$, from which on $t_*$ decreases until it reaches $0$. 

We first observe that unlike for $\lambda = 0$, $z(t_*)$ does not vanish as $t_* \rightarrow 0$, but it diverges for $\lambda = 1$. This shows that the correlator does not blow up except in the short distance limit $t_* \rightarrow t_0$. The pole in the two-point correlator occurring in the classical theory is therefore resolved. The dual field theory state can therefore be normalisable, unlike in the classical gravity limit \cite{EngelhardtFurtherHolographicInvestigations}. 

There still exists a clear signature of the classical pole in the form of a local minimum of $z(t_*)$ for $\lambda = 1$ around $t \approx 0.4$. The associated boundary separation is somewhat outside of the classical horizon scale, where the red curve intersects the x-axis. Taking $\lambda \rightarrow 0$, the value that the curve takes at its local minimum approaches $0$, but the characteristic turnaround behaviour persists except for $\lambda = 0$.

We also see that for $\lambda =1$, there exists a regime between the local maximum and minimum of $x(t_0)$, where three real geodesics with the same boundary separation exist. The two-point correlator is obtained by adding their contributions, in addition to complex solutions discussed in section \ref{sec:ComplexGeodesics}. 
If $t_0$ becomes close enough to $\lambda$, the characteristic behaviour shown in figure \ref{fig:PlotXZ} changes, as shown in figure \ref{fig:PlotXZt0}.

\subsection{Complex geodesics} \label{sec:ComplexGeodesics}

The importance of including complex geodesics was emphasised in \cite{EngelhardtHolographicSignaturesOf, EngelhardtFurtherHolographicInvestigations}. We will not add anything new to this topic but merely recall their results and comment on how complex geodesics influence the above results. 

First we note that the long distance contributions of complex geodesics to the two-point correlator turn out to be $\sim (\lbdy)^{-\frac{2 \Delta}{1-p}}$ for geodesics not crossing the singularity in the cases that have been studied in the classical setup \cite{EngelhardtHolographicSignaturesOf}. Our long range result \eqref{eq:RealContributionLR} from quantum corrected real geodesics is thus subdominant to the contributions from complex classical geodesics. These complex classical geodesics should be good approximations also in the context of the quantum corrected metric for $t_0 \gg \lambda$, since they run in regions of the spacetime which are well approximated by the classical theory. 

We have not investigated other complex geodesics so far.
Due to this, we cannot judge whether our results will be qualitatively affected by inclusion of possible complex solutions.
Already for reasons of continuity, one expects that such additional complex geodesics should be included.

\section{Comments} \label{sec:Comments}

A few comments are in order:

	In loop quantum gravity (and in the same way also in lattice gauge theory), it is necessary to approximate certain operators that are not well defined on the Hilbert space by functions of their exponentials, which introduces ambiguities in the quantisation. In the simplest homogeneous and isotropic case, the so called $\bar \mu$-scheme \cite{AshtekarQuantumNatureOf} leads to viable physical results and is also motivated from a full theory point of view by introducing significant quantum gravity corrections in the operator approximations only when the Planck curvature is reached \cite{BIII, BVI}. The prime example is to approximate $b \approx \frac{\sin (\lambda b)}{\lambda}$, where $b$ is the trace of the extrinsic curvature in spatially flat homogeneous and isotropic cosmology, i.e. derived from $ds^2_4$ in \eqref{eq:Metric4d5d}. 
	We have implicitly used this scheme tailored for 4d quantum gravity also in the 5d case. However, from a purely 5d point of view, it would be more appropriate to approximate $b \approx \frac{\sin (z \lambda b/L)}{z \lambda /L}$, since $b$ contains a Lie derivative w.r.t. the normal in $t$-direction, and $g_{tt} = -L^2/z^2$, as opposed to $-1$ in the 4d case. This leads to an effective $\lambda_{\text{5d}} = z \lambda_{\text{4d}}$ and
	\be
	 a(z,t) = \frac{\amin}{\lambda^{p}} \left(t^2 +  \frac{z^2 \lambda^2}{L^2} \right)^{p/2} \label{eq:QuantumMetricZ} \text{,}
	\ee
	where the $z$-scaling in the first $\lambda$ was absorbed in $\amin$. 
	Therefore, as one approaches the boundary, the quantum corrections would become negligible and the CFT background would be effectively classical. Also, the classical scaling symmetry \cite{EngelhardtHolographicSignaturesOf} $z \rightarrow \omega z, ~ t \rightarrow \omega t, ~ x_i \rightarrow \omega^{(1-p_i)} x_i$, which is broken for constant $\lambda$, holds in this case. 
	Conversely, for a $z$-independent $\lambda$ as in the main part of the paper, quantum effects appear at lower and lower 5d bulk scales as one approaches the boundary, however in the boundary theory they appear at the 4d Planck scale. \\
	Naively, evaluating the Christoffel symbols derived from \eqref{eq:QuantumMetricZ} at $z=0$ suggests that the singularity in the pole following from the classical computation is recovered even for the quantum corrected metric, since the null geodesic on the boundary responsible for it now exists, but this needs to be verified using a limiting procedure as in \cite{EngelhardtFurtherHolographicInvestigations}. In fact, our companion paper \cite{BodendorferHolographicSignaturesOf2} provides strong numerical evidence that this boundary geodesic is indeed isolated, thus leading to a resolution of the finite distance pole also for \eqref{eq:QuantumMetricZ}.

	In our computations, we have made the simplifying assumption that the geodesic equation still determines the propagation of test particles on the quantum corrected metric background. Within LQG, the validity of this assumption has not been established so far. In particular, recent work on the anomaly freedom of effective constraint algebras \cite{CailleteauAnomalyFreeScalar, BojowaldSomeImplicationsOf} suggests that additional significant quantum corrections might be necessary, but no consensus has been reached so far \cite{AchourANewLook}. Pioneering work on this question in the context of loop quantum cosmology has been done in \cite{AshtekarQuantumFieldTheoryOn}. 
	
	If it can be established from an LQG point of view that propagation happens along geodesics of the quantum corrected metric, or some other effective metric as discussed in \cite{AshtekarQuantumFieldTheoryOn}, this of course still doesn't mean that this would agree with a similar computation from string theory. Since, however, non-perturbative string theory is best defined via AdS/CFT, the relevant question to ask is again about the behaviour of the dual CFT. This suggests to simply check what the different proposals within LQG for field propagation on quantum geometry backgrounds yield for the dual CFT and to compare it with lattice simulations thereof. 
	
	Recent advances in pushing the Ryu-Takayanagi prescription \cite{RyuHolographicDerivationOf, HubenyACovariantHolographic} to the quantum regime \cite{FaulknerQuantumCorrectionsTo, EngelhardtQuantumExtremalSurfaces} have lead to many interesting insights into AdS/CFT beyond the classical supergravity approximation. It would be very interesting to compare these to a computation of minimal surfaces in quantum corrected backgrounds inspired by LQG. This might also suggest how to connect the present computations to full string theory.   

	In \cite{BIV}, a possible strategy to relate the large spin expansion in loop quantum gravity on a fixed graph to the $1/N$ expansion in AdS/CFT was discussed. In the current paper, this line of thought is not relevant as our quantum corrected solution can be seen to emerge from a minisuperspace quantisation, or can be embedded in a full theory context including a continuum limit in the case $p_i=1/3$. The value of $N^2$ as determined from the gravity side is given by the ratio of the AdS radius to the power 8 and the 10d Newton constant. In the present paper, we considered this ratio to be very large, as otherwise quantum corrections for the $z$-coordinate would have to be expected\footnote{It should be noted however that the quantum corrected metrics are not Ricci-flat any more in general \cite{SinghIsClassicalFlat}, which might provide a way to link them to finite $N$ effects.}. It is unclear to us at the moment to which extend our use of quantum gravity here can be linked to finite $N$ effects in the dual CFT. This also prevents us so far from making a comparison to $1/N$ corrections that have been computed using perturbative quantum supergravity, see e.g. \cite{Naculich1NCorrections, GubserDoubleTraceOperators, DenefBlackHoleDeterminants, Caron-HuotHydrodynamicLongTime, JorrinTowards1N}. 

	In \cite{EngelhardtFurtherHolographicInvestigations}, it is discussed that for a non-flat boundary metric, one obtains a singularity at $z=\infty$. This singularity can be removed by going over to 6 dimensions and slightly changing the metric. The same argument can be applied also in our case.

\section{Conclusion}  \label{sec:Conclusion}

In this paper, we have explored the holographic signature of a resolved cosmological singularity using a simple ansatz for the quantum corrected metric inspired by loop quantum gravity. It turned out that this ansatz gave sensible results for the dual CFT which can be seen as an improvement over the classical gravity approximation.  
The main open problem in this approach are additional complex solutions and their possible qualitative influence on the results. In addition, our possibility to perform analytic computations so far rests on using a regularisation tailored for 4d quantum gravity, as discussed in the comment section. This is unnatural from a 5d point of view as quantum effects appear on increasingly lower 5d scales as one approaches the boundary. This issue is tackled in a companion paper \cite{BodendorferHolographicSignaturesOf2}.

From a technical point of view, the relation of the results obtained using the geodesic approximation with the more established and general prescription of taking derivatives of the on shell action to compute the two-point function \cite{FreedmanCorrelationFunctionsIn} should be clarified, in particular its Lorentzian version \cite{SonMinkowskiSpaceCorrelators}. This may also be relevant to understand the fate of the finite distance pole in the two-point correlator for a 5d metric based on \eqref{eq:QuantumMetricZ}.
We leave this question to future work. 

Conceptually, our results are interesting since they give an example that AdS/CFT is the proper interface to ask the question of wether string theory, possibly in some limit, is related to other approaches to quantum gravity, for example loop quantum gravity. Our current results show that there is no obvious inconsistency. However, to answer this question reliably (with non-perturbative string theory defined via its dual CFT), we need to first understand the dual CFT better, for example using lattice methods, which is another long term goal of us. We hope that our rather naive and basic computation so far can serve as a basis for further explorations of these questions and stimulate some discussion. Merging ideas from AdS/CFT and loop quantum gravity has certainly attracted some interest recently. For a selection of the relevant works, see \cite{FreidelReconstructingAdSCFT, BTTVIII, DittrichIsingModelFrom, BonzomDualityBetweenSpin, Bonzom3DHolography, BIV, SmolinHolographicRelationsIn, HanLoopQuantumGravityExact, DittrichQuasiLocalHolographic1, DittrichQuasiLocalHolographic2}.

There are several obvious further directions to explore. Next to embeddings of cosmological spacetimes into AdS space, one could also consider black hole spacetimes. Here, significantly less is known from loop quantum gravity (see \cite{AshtekarBlackHoleEvaporation} for a general expectation), but a simple ansatz in the spirit of equation \eqref{eq:QuantumMetric} for the resolved singularity may show an interesting behaviour in the dual theory. Important guidance for constructing an explicit solution could be provided by \cite{ChamseddineNonsingularBlackHole}. Numerous AdS/CFT results in the classical gravity approximation such as \cite{KrausInsideTheHorizon, FidkowskiTheBlackHole} already exist and should provide a good starting point. Also, different bulk probes, such as minimal surfaces, are interesting to study, e.g. starting from \cite{EngelhardtEntanglementEntropyNear}. The problem of generic spacelike singularities may be approached by an appeal to the BKL conjecture \cite{BelinskiiOscillatoryApproachTo}, by which the relevance of the present results focussing on a single homogeneous Kasner patch may extended.

\section*{Acknowledgements}
NB was supported by a Feodor Lynen Return Fellowship of the Alexander von Humboldt-Foundation and during final improvements of this work by an International Junior Research Group grant of the Elite Network of Bavaria. Discussions with Martin Ammon and Johanna Erdmenger are gratefully acknowledged. We specifically thank Martin Ammon for useful comments on a draft of this paper.


\end{document}